\documentclass[superscriptaddress,aps,preprintnumbers,amsmath,showpacs,amssymb,prd,nofootinbib,reprint]{revtex4-1}
\usepackage{bm, color} 
\usepackage{amssymb,amsfonts,slashed,amsthm,amsmath,graphicx, soul}
\usepackage[caption=false]{subfig}
\usepackage{longtable}
\begin{document}

\def\WY#1{
\textcolor{blue}
{#1}}
\def\WYC#1{\textcolor{blue}{\bf { WY:} #1}}
\def\WYS#1{\textcolor{blue}{\sout{#1}}}
%
%


\renewcommand{\figurename}{Fig.}
\renewcommand{\tablename}{Table.}
\newcommand{\Slash}[1]{{\ooalign{\hfil#1\hfil\crcr\raise.167ex\hbox{/}}}}
\newcommand{\bra}[1]{ \langle {#1} | }
\newcommand{\ket}[1]{ | {#1} \rangle }
\newcommand{\beq}{\begin{equation}}  \newcommand{\eeq}{\end{equation}}
\newcommand{\bef}{\begin{figure}}  \newcommand{\eef}{\end{figure}}
\newcommand{\bec}{\begin{center}}  \newcommand{\eec}{\end{center}}
\newcommand{\laq}[1]{\label{eq:#1}}  
\newcommand{\dd}[1]{{d \o d{#1}}}
\newcommand{\Eq}[1]{Eq.~(\ref{eq:#1})}
\newcommand{\Eqs}[1]{Eqs.~(\ref{eq:#1})}
\newcommand{\eq}[1]{(\ref{eq:#1})}
\newcommand{\ab}[1]{\left|{#1}\right|}
\newcommand{\vev}[1]{ \left\langle {#1} \right\rangle }
\newcommand{\bs}[1]{ {\boldsymbol {#1}} }
\newcommand{\lac}[1]{\label{chap:#1}}
\newcommand{\SU}[1]{{\rm SU{#1} } }
\newcommand{\SO}[1]{{\rm SO{#1}} }
\def\({\left(}
\def\){\right)}
\def\dt{{d \o dt}}
\def\diag{\mathop{\rm diag}\nolimits}
\def\Spin{\mathop{\rm Spin}}
\def\O{\mathcal{O}}
\def\U{\mathop{\rm U}}
\def\Sp{\mathop{\rm Sp}}
\def\SL{\mathop{\rm SL}}
\def\tr{\mathop{\rm tr}}
\newcommand{\OR}{~{\rm or}~}
\newcommand{\AND}{~{\rm and}~}
\newcommand{\EV}{ {\rm \, eV} }
\newcommand{\KEV}{ {\rm \, keV} }
\newcommand{\MEV}{ {\rm \, MeV} }
\newcommand{\GEV}{ {\rm \, GeV} }
\newcommand{\TEV}{ {\rm \, TeV} }
\def\o{\over}
\def\a{\alpha}
\def\b{\beta}
\def\c{\varepsilon}
\def\d{\delta}
\def\e{\epsilon}
\def\f{\phi}
\def\g{\gamma}
\def\h{\theta}
\def\k{\kappa}
\def\l{\lambda}
\def\m{\mu}
\def\n{\nu}
\def\p{\psi}
\def\q{\partial}
\def\r{\rho}
\def\s{\sigma}
\def\t{\tau}
\def\u{\upsilon}
\def\w{\omega}
\def\x{\xi}
\def\y{\eta}
\def\z{\zeta}
\def\D{\Delta}
\def\G{\Gamma}
\def\F{\Phi}
\def\P{\Psi}
\def\S{\Sigma}
\def\me{\mathrm e}
\def\ol{\overline}
\def\tl{\tilde}
\def\*{\dagger}
\def\H{H_{\rm ubble}}
\bibliographystyle{JHEP}

\renewcommand\bibname{\Large References}

\preprint{TU-1220}

\vspace{1.0cm}

\title{First Result for Dark Matter Search by WINERED}

\author{Wen Yin}
\affiliation{ Department of Physics, Tokyo Metropolitan University, 
Minami-Osawa, Hachioji-shi, Tokyo 192-0397 Japan}
\affiliation{Department of Physics, Tohoku University, Sendai, Miyagi 980-8578, Japan }
 \author{Taiki Bessho}
 \affiliation{PhotoCross Co. Ltd., 17-203 Iwakura-Minami-Osagicho,
Sakyo-ku, Kyoto 606-0003, Japan} 
\author{Yuji Ikeda}
\affiliation{Laboratory of Infrared High-resolution Spectroscopy, Koyama Astronomical Observatory, Kyoto Sangyo {University}\, Motoyama, Kamigamo, Kita-ku, Kyoto 603-8555, Japan} 
\affiliation{PhotoCross Co. Ltd., 17-203 Iwakura-Minami-Osagicho,
Sakyo-ku, Kyoto 606-0003, Japan} 
\author{Hitomi Kobayashi}
 \affiliation{PhotoCross Co. Ltd., 17-203 Iwakura-Minami-Osagicho,
Sakyo-ku, Kyoto 606-0003, Japan} 
\author{Daisuke Taniguchi}
\affiliation{National Astronomical Observatory of Japan, 2-21-1 Osawa, Mitaka, Tokyo 181-8588, Japan}
\author{Hiroaki Sameshima}
\affiliation{Institute of Astronomy, the University of Tokyo, 2-21-1 Osawa, Mitaka, Tokyo 181-0015, Japan}
\author{Noriyuki Matsunaga}
\affiliation{Department of Astronomy, Graduate School of Science, University of Tokyo, 7-3-1 Hongo, Bunkyo-ku, Tokyo 113-0033, Japan}
\author{Shogo Otsubo}
\affiliation{Laboratory of Infrared High-resolution Spectroscopy, Koyama Astronomical Observatory, Kyoto Sangyo {University}\, Motoyama, Kamigamo, Kita-ku, Kyoto 603-8555, Japan} 
\author{Yuki Sarugaku}
\affiliation{Laboratory of Infrared High-resolution Spectroscopy, Koyama Astronomical Observatory, Kyoto Sangyo {University}\, Motoyama, Kamigamo, Kita-ku, Kyoto 603-8555, Japan} 
\author{Tomomi Takeuchi}
\affiliation{Laboratory of Infrared High-resolution Spectroscopy, Koyama Astronomical Observatory, Kyoto Sangyo {University}\, Motoyama, Kamigamo, Kita-ku, Kyoto 603-8555, Japan} 
\author{Haruki Kato}
\affiliation{Laboratory of Infrared High-resolution Spectroscopy, Koyama Astronomical Observatory, Kyoto Sangyo {University}\, Motoyama, Kamigamo, Kita-ku, Kyoto 603-8555, Japan} 
\author{Satoshi Hamano}
\affiliation{National Astronomical Observatory of Japan, 2-21-1 Osawa, Mitaka, Tokyo 181-8588, Japan}
\author{Hideyo Kawakita}
\affiliation{Laboratory of Infrared High-resolution Spectroscopy, Koyama Astronomical Observatory, Kyoto Sangyo {University}\, Motoyama, Kamigamo, Kita-ku, Kyoto 603-8555, Japan} 
\affiliation{Department of Astrophysics and Atmospheric Sciences, Faculty of Science, Kyoto Sangyo University}

\begin{abstract}
The identity of dark matter has been a mystery in astronomy, cosmology, and particle theory for about a century. We present the first dark matter {search with a high-dispersion spectrograph} by using WINERED at $6.5$m Magellan Clay telescope to measure the photons from the dark matter decays. 
The dwarf spheroidal galaxies (dSphs) Leo V and Tucana II are observed by utilizing an object-sky-object nodding observation technique.
 Employing zero consistent flux data after the sky subtraction and performing Doppler shift analysis for further background subtraction, we have established one of the most stringent limits to date on dark matter lifetime in the mass range of $1.8-2.7\,$eV. The conservative bound is translated to the photon coupling, $g_{\phi\gamma\gamma}$ for axion-like particles around $g_{\phi\gamma\gamma}\lesssim (2-3)\times 10^{-11}\,$GeV$^{-1}$ ($10^{-10}\,$GeV$^{-1}$) for the case that ultra-faint dSphs have the Navarro-Frenk-White (generalized Hernquist) dark matter profile.
\end{abstract}

\maketitle
\flushbottom

\vspace{0.2cm}
{\bf Introduction.--}
Dark Matter has played a crucial role in the evolution of the universe, and its existence is now considered certain in the current universe. 
Yet, despite numerous experimental searches, the basic properties of dark matter, such as mass and non-gravitational interactions, have remained unknown for about a century. The spatial distribution of dark matter within galaxies is typically inferred from gravitational interactions, evident in phenomena like the rotation curves of galaxies.

One candidate for dark matter is an undiscovered elementary particle known as the axion-like particle (ALP), which couples with photons. 
Theoretically, scenarios like the ``ALP miracle"\cite{Daido:2017wwb, Daido:2017tbr} (see {Refs.\,\cite{IAXO:2019mpb, Daido:2017tbr} for the detailed parameter region and  \,\cite{Takahashi:2023vhv} for the scenario that the strong CP problem can be explained) predicts the dark matter in the mass range of 0.01-7.7 eV with a particular mass-photon coupling relation. 
 In this scenario, the ALP gluon coupling is suppressed, and various constraints from e.g. thermal production and astrophysical productions are alleviated c.f.~Refs.\,\cite{Chang:1993gm, Moroi:1998qs, Caloni:2022uya, Lella:2023bfb}. 
 It predicts the existence of ALP with masses near eV, decaying into two photons, producing narrow-line emissions in the infrared region.}
{Alternatively, the hot dark matter paradigm, prevalent about 40 years ago, suggests a dark matter mass around the eV scale. Although dark matter produced thermally tends to have too long a free-streaming length, it can be cold dark matter if the Bose-enhancement effect is significant in thermal production. This is because stimulated emission is IR dominant and reaches a steady state that the small momentum modes are populated~\cite{Yin:2023jjj}.\footnote{{We also note that there are many production mechanisms that can produce dark matter in but are not restricted to, around the eV mass range: e.g., although the minimal vacuum misalignment mechanism does not apply to the parameter region of our interest~\cite{Preskill:1982cy,Abbott:1982af,Dine:1982ah}, one may introduce dark sector topological susceptibility~\cite{Arias:2012az} and a flat-top-shaped potential with a hilltop condition~\cite{Nakagawa:2020eeg} to produce the dark matter in the region of interest. }} 
Phenomenologically, when the photon coupling of the ALP dark matter is $g_{\phi\gamma\gamma}= \O(10^{-10})\,\mathrm{GeV}^{-1}$, it can explain two independent excesses  relevant to the cosmic infrared background simultaneously if the mass is around $2\,\mathrm{eV}$~\cite{Caputo:2020msf,Korochkin:2019qpe,Bernal:2022xyi} (see also Refs.\,\cite{Gong:2015hke, Kohri:2017oqn, Kalashev:2018bra, Kashlinsky:2018mnu, Nakayama:2022jza, Carenza:2023qxh}).} 

{Recently, motivated by those non-trivial coincidences, authors in Ref.\,\cite{Bessho:2022yyu} proposed that the state-of-the-art infrared spectrographs, such as Warm INfrared Echelle spectrograph for Realizing Extreme Dispersion and sensitivity (WINERED) installed at the $6.5$m Magellan Clay telescope and the Near-Infrared Spectrograph (NIRSpec) on the James Webb Space Telescope (JWST), can efficiently search for {the }eV range dark matter.  This is particularly relevant if the dark matter decays into a photon with a narrow line spectrum, as background noises usually constitute continuous spectra.}

In this paper, we {use} the WINERED~\cite{ikeda2006winered, yasui2008warm, kondo2015warm, ikeda2016high, ikeda2018very, 2022WINERED}, a near-infrared, high-resolution spectrograph {developed} by the University of Tokyo alongside the Laboratory of Infrared High-resolution Spectroscopy at Kyoto Sangyo University, to search for dark matter.
It is a PI-type instrument that began operating with the $3.58$m New Technology Telescope (NTT) at La Silla Observatory in 2017 and was subsequently fitted to the $6.5$m Magellan Clay telescope at Las Campanas Observatory. Offering unparalleled sensitivity with an instrumental throughput reaching up to $50\%$, WINERED stands out among {near-infrared high-resolution spectrographs}, attachable to telescopes ranging from intermediate (3-4m class) to large (8-10m class), especially in the short NIR spectrum (0.9-1.35$\mu$m). 
WINERED is equipped with three observational settings: the “WIDE” mode ($R=28000$), the “HIRES-Y” mode ($R=68000$), and the “HIRES-J” mode ($R=68000$). Here $R\equiv \l/\D \l_{\rm FWHM}$ quantify the spectral resolution with $\l$ ($\D \l_{\rm FWHM}$) being the 
wavelength (peak full width at half of the maximum hight).

On July 6th, 2023, we used the ``WIDE" mode of WINERED on the Magellan Telescope to perform 1 hour observation of the Leo V dwarf spheroidal galaxy (dSph), 
and 0.5 hour observation for the blank sky to subtract the background noise to search for dark matter. 
We found some excesses in the data. One way to further reduce the noise is to examine the Doppler shift of the wavelengths, which arises from the varying radial velocities among different dSphs as suggested in~\cite{Bessho:2022yyu}. {Specifically, we will check if the potential intrinsic sources of the line spectra are at rest in any dSphs.}  
Therefore, we performed the second observation on November 2nd, 2023, by looking at the Tucana II for 1.2 hours with a blank sky observation of 0.7 hours.

In this paper, we use data consistent with zero after sky subtraction to establish a {limit} on the dark matter, assuming that dark matter decays into two particles, one of which is a photon. 
We find that our observational data can set one of the strongest {limit}s on dark matter decay rate in the mass range of $1.8-2.7\EV$.
Our {result} is robust because we do not rely on a specific background model. 

{Here, we mention the dark matter search using the NIRSpec at JWST. Although it was discussed that interstellar absorption makes the search around the Milky Way galactic center less effective compared to searching for dark matter in certain dSphs~\cite{Bessho:2022yyu}, the search of non-galactic center was performed by analyzing public blank sky data by subtracting continuous spectra~\cite{Janish:2023kvi}. This turned out to be an economical approach without proposing a specific observation, and the result is adjusted by including the effects of interstellar absorption and different choices of the $D$-factor (see \Eq{D-factor} for definition)~\cite{Roy:2023omw}.
 The future reach was also estimated in Refs.~\cite{Bessho:2022yyu,Janish:2023kvi,Roy:2023omw}. In the optical range, indirect detection of dark matter was also performed~\cite{Grin:2006aw,Regis:2020fhw,Todarello:2023hdk}. Since the observational targets, dark matter mass ranges, and methods are different, the constraints are complementary. For instance, our analysis
can also constrain spectra other than line ones (e.g. Ref.~\cite{Jaeckel:2021ert}), while the analysis subtracting continuous spectra cannot. 
}

{\bf Lines from Dark Matter in dSphs.--}  
Our main focus in this paper is the dark matter that decays into two particles, one of which is a photon. 
Here, we define the decay rate as $\Gamma_{\f}$, which also represents the inverse of the lifetime, and denote the number of photons in the final states by $q$. The mass of the dark matter is $m_\phi$. 
We assume that both daughter particles are significantly lighter than the dark matter and thus neglect their masses. For instance, the well-studied ALP dark matter has
$
\Gamma_\f= \frac{g_{\f\g\g}^2}{64\pi} m_\f^3,
$
where $g_{\f\g\g}$ is the ALP coupling to photons, and $q=2$.

 The photon differential flux at the outer surface of the Earth's atmosphere from this decay comprises two parts,
\begin{align}
\frac{d\Phi_\g}{d E_\g}= \frac{d\Phi_\g^{\text{extra}}}{d E_\g}+\sum_i\frac{d\Phi_{\g,i}}{d E_\g}
\end{align}
where the {first term} on the right-hand side represents an isotropic extragalactic component with a characteristic energy scaling. We do not consider this component in our analysis as it would induce a continuous spectrum that gives {lower} sensitivity compared to the {second} term on the right-hand side.
In addition, it will be reduced by the sky subtraction in our analysis.  The second term is our focal point, {and we consider} the emission from a given nearby galaxy $i$ in the angular direction of $\Omega$. Angular differential flux can be expressed as
\begin{align}
\frac{\partial ^2\Phi_{\gamma,i}}{\partial\Omega \partial E }
&=\int ds  \frac{e^{-\tau[s, \Omega ] s}}{4\pi s^2} \left(\frac{ q\Gamma_{\phi} \rho^{i}_\phi(s, \Omega )}{m_\phi}\right) s^2 \frac{\partial^2 N_{\phi,i}}{q \partial E\partial \Omega}[ s, \Omega] \nonumber\\ 
&\simeq \frac{\partial_\Omega D}{4\pi} \frac{q \Gamma_\phi}{m_\f} \frac{\partial^2 {N}_{\phi,i}}{q \partial E\partial \Omega}.
\end{align}
In this equation, $s$ denotes the line-of-sight distance.
$\rho^{i}_\phi$ and $\frac{\partial^2 N_{\phi,i}}{q \partial E\partial \Omega}[ s, \Omega]$ signify the dark matter density profile and the emission spectrum of each photon from an individual dark matter decay in the vicinity of galaxy $i$, respectively. 
Both of the factors are contingent upon the dark matter profile models and the attributes of galaxy $i$ in general.
In the approximation, we assume that the dependence on $s$ and $\Omega$ in $\partial^{2} N_{\f,i}/\partial E\partial \Omega$ is not significant, and we factorize the integral. 
Indeed, those decay signals from dSphs look like line spectra for the WINERED in the WIDE mode, in which $R= 28000$, given that the typical velocity dispersion of the dark matter {around the center of a dSph is much smaller than 10km/s~\cite{Bessho:2022yyu}}. 
{Then, by neglecting the velocity dispersion of the dark matter we can approximate}
$
\frac{\partial^2 N_{\phi, i} }{q \partial E \partial \Omega} =2 \delta{\left(E-\frac{m_\phi}{2} (1-v_{ i})\right)}.  
$
{Here and hereafter, we use the natural units}.
{Note that we cannot neglect the radial velocity, $v_i$, of the dSph since it is typically much larger than $1/R\approx 3.57\times 10^{-5}$. }
For instance, the {heliocentric} radial velocities of Leo V and Tucana II are~\cite{Wenger:2000sw}, 
$
v_{\rm LeoV}\approx 5.78\times 10^{-4}, \AND v_{\rm TucII}\approx -4.31\times 10^{-4},
$
respectively.
The term 
\begin{equation}\laq{D-factor} 
\partial_\Omega D_i\equiv \int ds \rho_\f^{i} 
\end{equation} 
is the so-called differential $D$-factor \cite{Combet:2012tt,Geringer-Sameth:2014yza,Bonnivard:2015xpq,Bonnivard:2015tta,Hayashi:2016kcy,Sanders:2016eie,Evans:2016xwx,Hayashi:2018uop,Petac:2018gue,Salucci:2018hqu},  which is carefully derived in \cite{Yin:2023uwf} as a function of the angle $\Omega$ measured from the center of galaxy (see also \cite{Hayashi:2020jze,Hayashi:2022wnw}). For Leo V, $\log_{10}{\(\partial_\Omega D_{\rm LeoV}/[\rm GeV/cm^2/sr]\)}\approx 21.4^{+0.9}_{-0.7}$ around the center of the galaxy in the level of $\O({\rm arcsec})$, 
{which will be the position we look at.}
 For Tucana II, $\log_{10}{\(\partial_\Omega D_{\rm TucII}/[\rm GeV/cm^2/sr]\)}\approx 22.2^{+1.4}_{-0.9}$. Here, we adopted the estimation in Ref.\,\cite{Yin:2023uwf}, {{by using the latest observational data for the dSphs},  by utilizing a generalized Hernquist profile~\cite{1990ApJ...356..359H,1996MNRAS.278..488Z}, which takes account of nonspherical mass distributions and the possibilities of cored and cuspy dark matter profiles, fitted to observational data.}
$\tau $ represents the (averaged) optical depth. 
For the targets and wavelengths of interest, 
we can neglect the scattering or absorption of the photon during the propagation to {the outer surface of Earth's atmosphere}. 
However, those effects due to the atmosphere of Earth {cannot be neglected}. To estimate the flux on ground-based telescopes, we {simply use} 
$
\eta[2\pi/E_\g] \times \frac{\partial ^2\Phi_{\gamma,i}}{\partial\Omega \partial E_\gamma }|_{\tau=0},
$
where $\eta[\l]$ denotes the atmospheric transmittance{, which we will measure from the observation of a standard star by using the software package Molecfit~\cite{smette2015molecfit,kausch2015molecfit}.}

From those discussions and by applying the estimate in Ref.\,\cite{Yin:2023uwf},
$
{\partial_\Omega\Phi_{\gamma,\rm Leo V}}\approx 1.33^{+8.28}_{-1.07}\times 10^{-6} {\rm cm^{-2}s^{-1} arcsec^{-2}}\(\frac{2\EV}{m_\f}\)\frac{q\Gamma_{\f}}{\(1.65\times 10^{24}\rm s\)^{-1}},
{\partial_\Omega\Phi_{\gamma,\rm  TucII}}\approx 9.83^{+210.14}_{-8.59}\times 10^{-6} {\rm cm^{-2}s^{-1} arcsec^{-2}}\(\frac{2\EV}{m_\f}\)\frac{q\Gamma_{\f}}{\(1.65\times 10^{24}\rm s\)^{-1}}.
$
The flux will be compared with the observational data to set a {limit} on the dark matter decay rate.

\begin{table*}[t]
\caption{Obervation logs. Here, Regions 1, 2, and 3 are for Leo V, Tucana II, and Tucana II, respectively.
$R$ is the spectral resolution. $T_I$ denotes the total integration time. 
}
\label{table:obs_log}
\centering
\begin{tabular}{llcccccc}
\hline
Object name &Object type &RA(J2000) & DEC(J2000) & Obs. date & $J_m$ & $R$ &$T_I$ (sec) \\
\hline \hline
Leo V & dSph & 11:31:09.6 & +02:13:12 & 2023.06.06 & $-$ & 28,000 & 3600 \\
Tucana II & dSph & 22:51:55.1 & -58:34:08 & 2023.11.02 & $-$ & 28,000 & 4200 \\
Sky region 1 
& $-$ & 11:31:56.97 & +02:09:19 & 2023.06.06 & $-$ & 28.000 &1800 \\
Sky region 2 
& $-$
& 22:51:06.5 & -57:28:46 & 2023.11.02 & $-$ & 28,000 & 1200 \\
Sky region 3 
& 
{$-$}
& 22:38:08.1 & -58:24:39 & 2023.11.02 & $-$ & 28,000 & 1200 \\
HD134936 & A0V & 15:14:41.4 & -52:35:42& 2023.06.06 & 9.44 & 28,000 & 90 \\
\hline
\end{tabular}
\end{table*}

{\bf Observation.--} Observations were conducted using the WINERED \cite{2022PASP134a5004I}
attached to the 6.5m Magellan Clay telescope on July 6th and November 2nd in 2023.
We used the WIDE mode with a 0.29 arcsec slit {that allows us to cover the}
 wavelength region of $\lambda = 0.92 -1.35$ $\mu$m with
a spectral resolution of $R=28,000$.
The observation targets were Leo V (on 6th, Jul, 2023) and Tucana II (on 2nd, Nov, 2023), 
which are dSphs located at a distance of 178 kpc and 57 kpc, respectively.
The spectrometer slit (0.29 arcsec $\times$ 8.6 arcsec) was positioned
near the optical centroids of dSphs.
At that time, we took care to ensure that members of dSphs or foreground stars did not enter
the slit using WINERED's slit viewer, which is sensitive to the $R$-band
($\lambda_0 = 0.659$ $\mu$m) and $I$-band ($\lambda_{0} = 0.800$ $\mu$m).
The integration time per frame was set to 900 sec for Leo V and 600 sec for Tucana II.
The maximum integration time per frame was determined to minimize the effects of
(i) saturation from telluric night airglow and
(ii) spectral-shifting on the detector due to changes in ambient temperature
(Since the slight spectral shifting appeared on the observation in July, we reduced the maximum integration time
to 600 sec for the observation in November.)
Four frames of Leo V and seven frames of Tucana II were taken with a total exposure time of 3600 sec and 4200 sec,
respectively.
In order to subtract the components of the telluric lines,
and {ambient background radiation, such as zodiacal light,}
sky frames were obtained with the same integration time for each target.
The sky regions (one region for Leo V and two regions for Tucana II) were
selected to be more than 15 arcmin away from targets in areas without known infrared point sources.
Furthermore, for the absolute flux calibration, HD134936 (A0V, $J_m = 9.44$) was observed
as the photometric standard star.
A0V stars have only {a few strong intrinsic absorption lines} from the photosphere
in the WINERED wavelength region, making them suitable for use as photometric standard stars
in high-resolution spectroscopy. The spectrum of HD134936 was used not only
for the flux calibration of Leo V observed on the same night but also for that of Tucana II on a different night. {The atmosphere transmittance, $\eta(\l)$, is also measured.} On the night of November 2nd, the wind was extremely strong, and
due to the observatory's regulations, we were forced to interrupt the observations midway.
As a result, we were unable to acquire spectra of the standard star on the Nobemver's run.
However, it is known that the efficiency of the WINERED instrument is very stable,
and even including variations in atmospheric transmittance, the impact of using a
standard star from different nights is not very significant, especially for the wavelength regime $\eta(\l)\approx1$ {(see, e.g., Ref.\,\cite{1998PASP110200F})}.
The observation log is shown in Table \ref{table:obs_log}.

{\bf Data reduction.--}
For the data analysis, we utilized the WINERED Automatic Reduction Pipeline (WARP, \cite{Hamano_2024}),
provided as the standard pipeline for WINERED.
WARP is capable of executing the basic analysis necessary for spectra obtained by WINERED,
including sky/scattered light subtraction, flat fielding, interpolation of cosmic-ray-contaminated/bad pixels,
transformation of two-dimensional spectra, and wavelength calibration.
However, since we found a wavelength shift of approximately 0.1 pixels
during a single exposure ($t=900$ sec) in the raw data of Leo V
(even such a small shift can introduce significant systematic errors after sky subtraction
due to the peaky profile of airglow lines), we prepared corrected two-dimensional images for the wavelength shifts beforehand and input these data sets into WARP for analysis.
By summing up the two-dimensional echelle spectra along the spatial direction (along the slit length)
for each order produced by WARP, we finally obtained the average spectrum $F(\lambda)$ and error spectrum $\sigma(\lambda)$
of the central regions of targets. {They are defined in the units of ${\rm  cm^{-2} s^{-1} sr^{-1}} $.}
In conducting flux calibration for the slit spectrometer,
it was necessary to estimate the amount of slit loss due to the standard star's image size being wider than the slit width.
We estimated the slit loss from the assumption that the stellar image is a Gaussian profile and the seeing size (FWHM)
was measured from the spatial-direction cross-section of the two-dimensional spectrum of the standard star on the detector.

{\bf Spectrum-independent {analysis}.--}
We can set a $2\sigma$ {limit} on the decay rate of dark matter by requiring the dark matter flux to be smaller than the reduced spectra in the $2\s$ range, i.e. $\partial_\Omega \F_{\g,i}<2 \sigma_i[\frac{4\pi }{m_\phi (1-v_i)} ]/\eta[\frac{4\pi }{m_\phi (1-v_i)} ].$
In our analysis, we exclusively utilized data where $|F(\lambda)|< 2\sigma(\lambda)$, excluding data points where $|F(\lambda)|> 2\sigma(\lambda)$. 
This threshold ensures that we consider only data consistent with statistical fluctuations around zero following sky subtraction. 
To synthesize the data, we selected the most stringent limits among those derived from the data of Tucana II and Leo V for each bin.
Each bin is defined by the range 
$[m_\f(1-1/28000)^{-1/2}$-$m_\f(1-1/28000)^{1/2}]$. We varied $m_\f$ discretely to cover the whole range from 1.8eV to 2.7eV without overlapping. 
{We include the effect of the Doppler shifts due to the radial velocities of the dSphs (by taking account of the rotation and orbital motion of the Earth). Thus our analysis is done on the dSph frames. Therefore the intrinsic line emission/absorption on the Earth frame is further reduced.}
The combined results are given in the supplementary material. 
In the analysis so far, we did not assume the signal to be a line spectrum, and the data can be useful for various other scenarios.  

{\bf Results.--}  To constrain line spectra, we perform a subtraction using a spline fit {over each Echelle order by using 20-50 points}.
This method helps to eliminate some systematic errors, particularly under the assumption that the signal conforms to a line spectrum.
It allows us to increase the number of bins with photon flux consistent with zero.
The resulting {limit} on the decay rate is illustrated in the red band in the upper panel of Fig.\,\ref{fig:const2} by taking account of the uncertainty in the $D$-factor, {and the gray shaded region is excluded even with the conservative D-factor in the generalized Hernquist profile in the cored-like profile limit in the ultra-faint dSphs \cite{Yin:2023uwf}. However, according to the recent $N$-body and hydrodynamical simulations (e.g., \cite{Tollet:2015gqa,Lazar:2020pjs}), the conventional Navarro-Frenk-White(NFW) profile \cite{Navarro:1995iw, 2010MNRAS.402...21N, Burkert:1995yz} is predicted.  {This could be understood from suppressed baryonic feedback in the ultra-faint dSphs.} Thus, we also employ the differential $D$-factor for the NFW profile~\cite{Evans:2016xwx},  
 $\log_{10}{\(\partial_\Omega D^{\rm NFW}_{\rm LeoV}/[\rm GeV cm^{-2}sr^{-1}]\)}\approx 22.4^{+0.5}_{-0.4}$ and $\log_{10}{\(\partial_\Omega D^{\rm NFW}_{\rm TucII}/[\rm GeV cm^{-2}sr^{-1}]\)}\approx 22.9^{+0.4}_{-0.3}$ for setting the limit in the blue band for comparison~(see supplemental material for the derivation). The above the blue band is excluded with using the smallest values in the uncertainty.
}

{The lower panel of the figure is the same one translated into the photon coupling of the ALP dark matter with $q=2$.\footnote{{20 peaks in the figure are due to the efficiency of WINERED, i.e.\ the boundary effect of the slits for the echelle gratings}. }
For comparison, the bounds from the globular clusters (the horizontal dashed line)~\cite{Dolan:2022kul,Ayala:2014pea}, JWST blank sky observations adopting NFW profile of the Milky Way galaxy~\cite{Roy:2023omw,Janish:2023kvi}, and MUSE indirect detection  adopting NFW profiles of various dSphs (in the right region)~\cite{Todarello:2023hdk} (See Ref.\,\cite{AxionLimits} for some data we used) are also shown together with the ALP miracle prediction~\cite{Daido:2017wwb,Daido:2017tbr,IAXO:2019mpb}(see Ref.\,\cite{Bessho:2022yyu} for the plot).}
Our observations and analysis enable us to set one of the strongest {limit}s
 to date on dark matter decaying into a photon with a line spectrum in the mass range of  $1.8-2.7\EV$.

\begin{figure}[!t]
   \includegraphics[width=65mm]{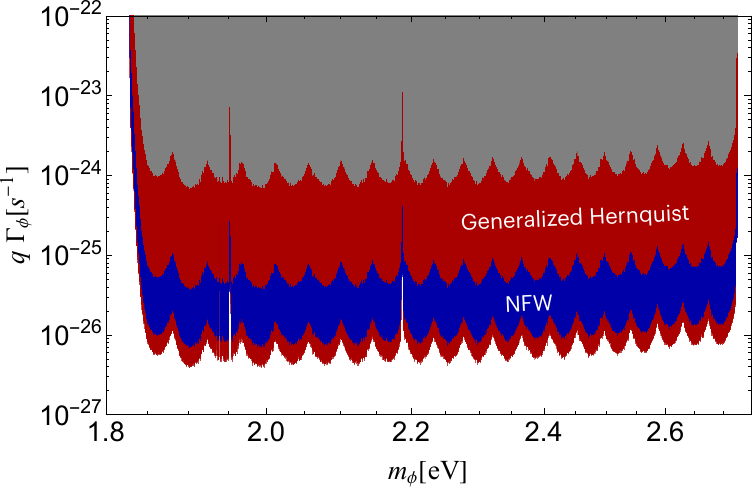}
   \\
   \includegraphics[width=70mm]{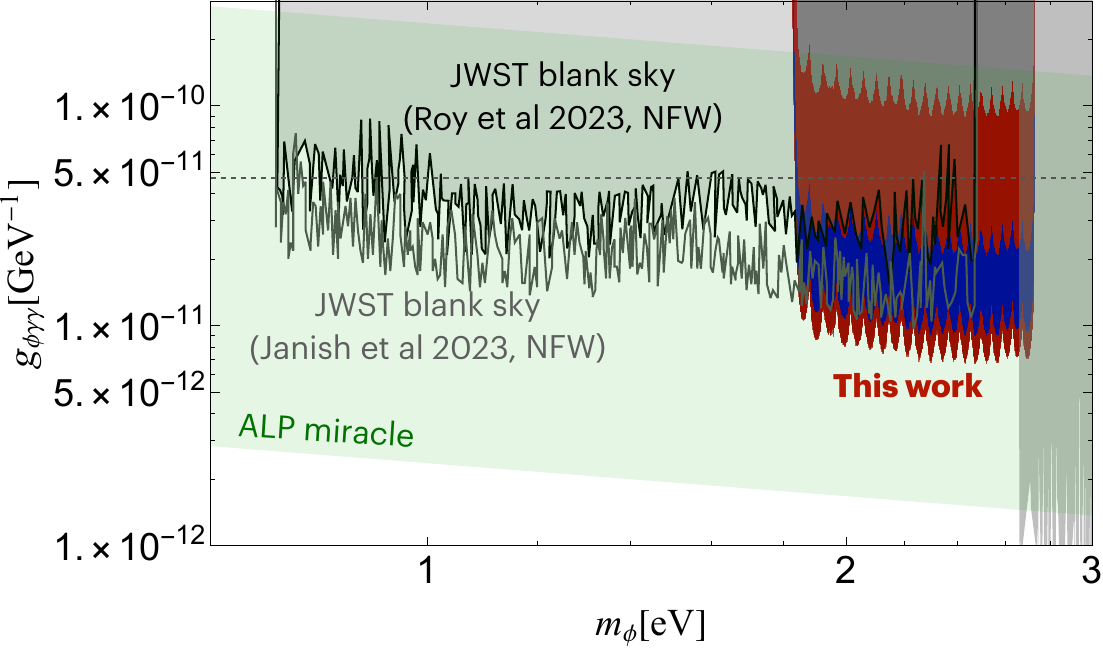}
      \vspace{-5mm}
\caption{{We show the $2\sigma$ limit for dark matter decaying into two massless particles, one of which is a photon. The top panels show the constraints on the decay rate multiplied by the number of photons in the final state, $q$, as a function of the mass $m_\phi$. The red band (blue band) corresponds to the uncertainty in the differential $D$-factor from the generalized Hernquist profile (the NFW profile). The gray region is excluded even in the generalized Hernquist profile case, while above the blue band is excluded in the NFW profile case by assuming conservative $D$-factors. The bottom panel displays the translated constraints on the photon coupling, $g_{\phi\gamma\gamma}$, for the ALP dark matter ($q=2$). The horizontal dashed line denotes the bound from stellar cooling. The limits from JWST blank sky observations by Refs.~\cite{Janish:2023kvi,Roy:2023omw} and MUSE~\cite{Todarello:2023hdk} are also shown for comparison.}
      \vspace{-5mm}
} \label{fig:const2}
\end{figure}

{\bf Discussion.--}
We note that $\O(1)\%$ of the total bins remain unconstrained, since the condition, $|F_{i}(\frac{4\pi}{m_\f (1-v_i)})|>2\s_{i}(\frac{4\pi}{m_\f (1-v_i)})$ holds both for Tucana II and LeoV. They are, therefore, not included in the figures, although they are difficult to see due to the high resolution. In this sense, we did not completely exclude the region that can be seen.
These data points will be crucial in our future studies.
In the data that were not included in the present analysis, we also found excesses of more than 5 sigma local significance by performing a fit by a narrow Gaussian line combined with a smooth spline function. 
Interestingly, a few of these excesses overlap in the dSphs' frames.
However, it must be noted that the observational conditions were not ideal. 
Therefore, while we observe intriguing excesses, we are cautious about claiming the detection of intrinsic line signals at this stage. 
Nonetheless, these findings underscore the importance of conducting further observations.

\begin{acknowledgments}
W.Y. would like to thank K. Hayashi for the useful discussion on the dark matter profiles. 
This work was supported by JSPS KAKENHI Grant Nos.  22K14029 (W.Y.), 20H05851 (W.Y.), 21K20364 (W.Y.), and 22H01215 (W.Y.) and Incentive Research Fund for Young Researchers from Tokyo Metropolitan University (W.Y.).
This paper is based on WINERED data gathered with the 6.5
m Magellan Clay Telescope located at Las Campanas
Observatory, Chile, under the proposal ``eV-Dark Matter search with WINERED"
(PI: Wen Yin, Co-Is: Yuji Ikeda and Taiki Bessho).
We thank the staff of Las Campanas Observatory for their support during
the WINERED's installation and observations. WINERED was
developed by the University of Tokyo and the Laboratory of
Infrared High-resolution Spectroscopy, Kyoto Sangyo University,
under the financial support of KAKENHI (Nos.
16684001, 20340042, and 21840052) and the MEXT Supported
Program for the Strategic Research Foundation at
Private Universities (Nos. S0801061 and S1411028). The observing runs in 2023 June and November were partly supported by KAKENHI (grant No. 19KK0080),
JSPS Bilateral Program Number JPJSBP120239909, and the Project Research Number AB0518
from the Astrobiology Center, NINS, Japan. 
\end{acknowledgments}

\appendix
\clearpage
\maketitle
\onecolumngrid
\begin{center}
\textbf{\large First Result for Dark Matter Search by WINERED}\\
\vspace{0.05in}
{ \it \large Supplementary Material}\\ 
\vspace{0.05in}
\end{center}

\subsection{Slit images}
\begin{figure}[!h]
\begin{center}  
   \includegraphics[width=65mm]{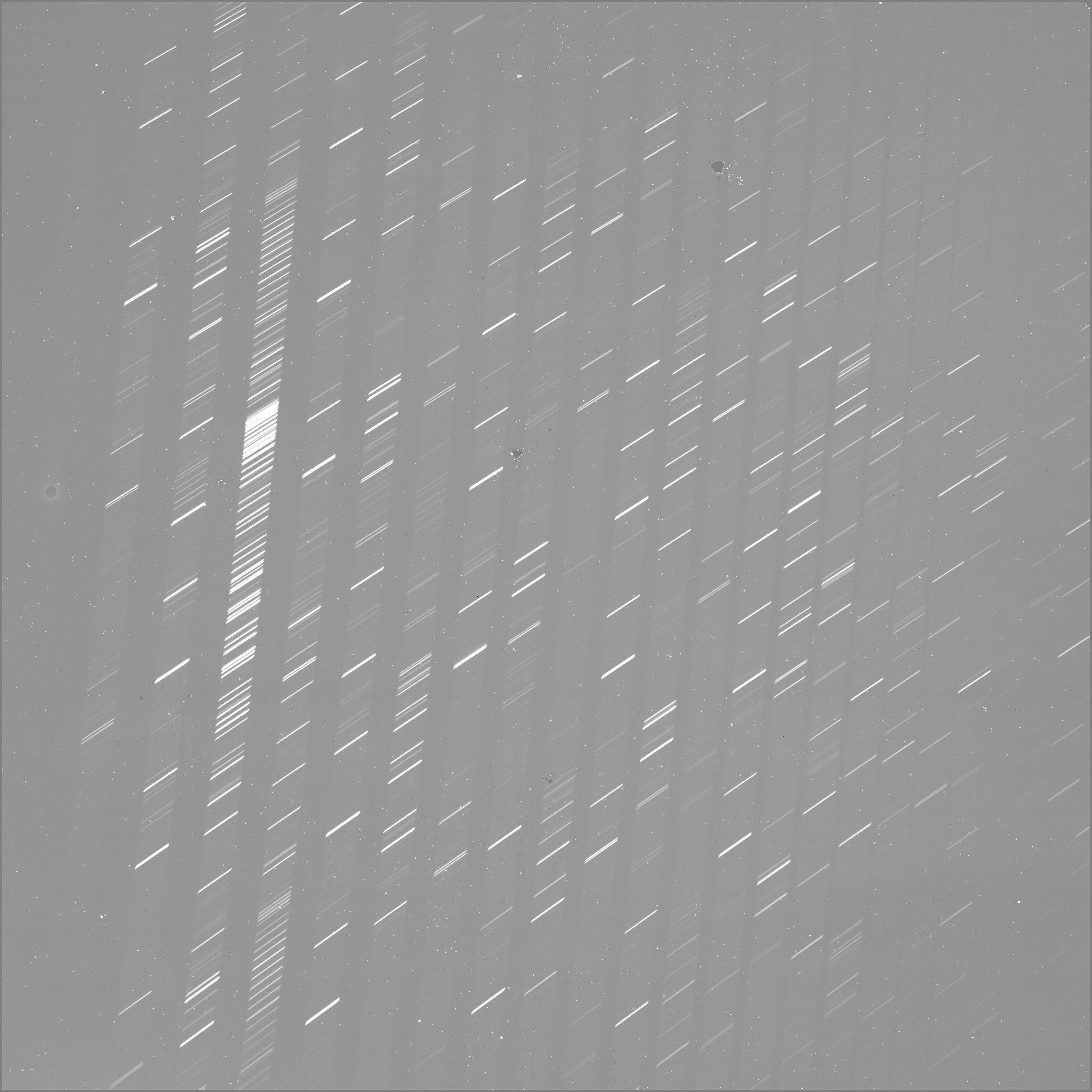}  
    \includegraphics[width=65mm]{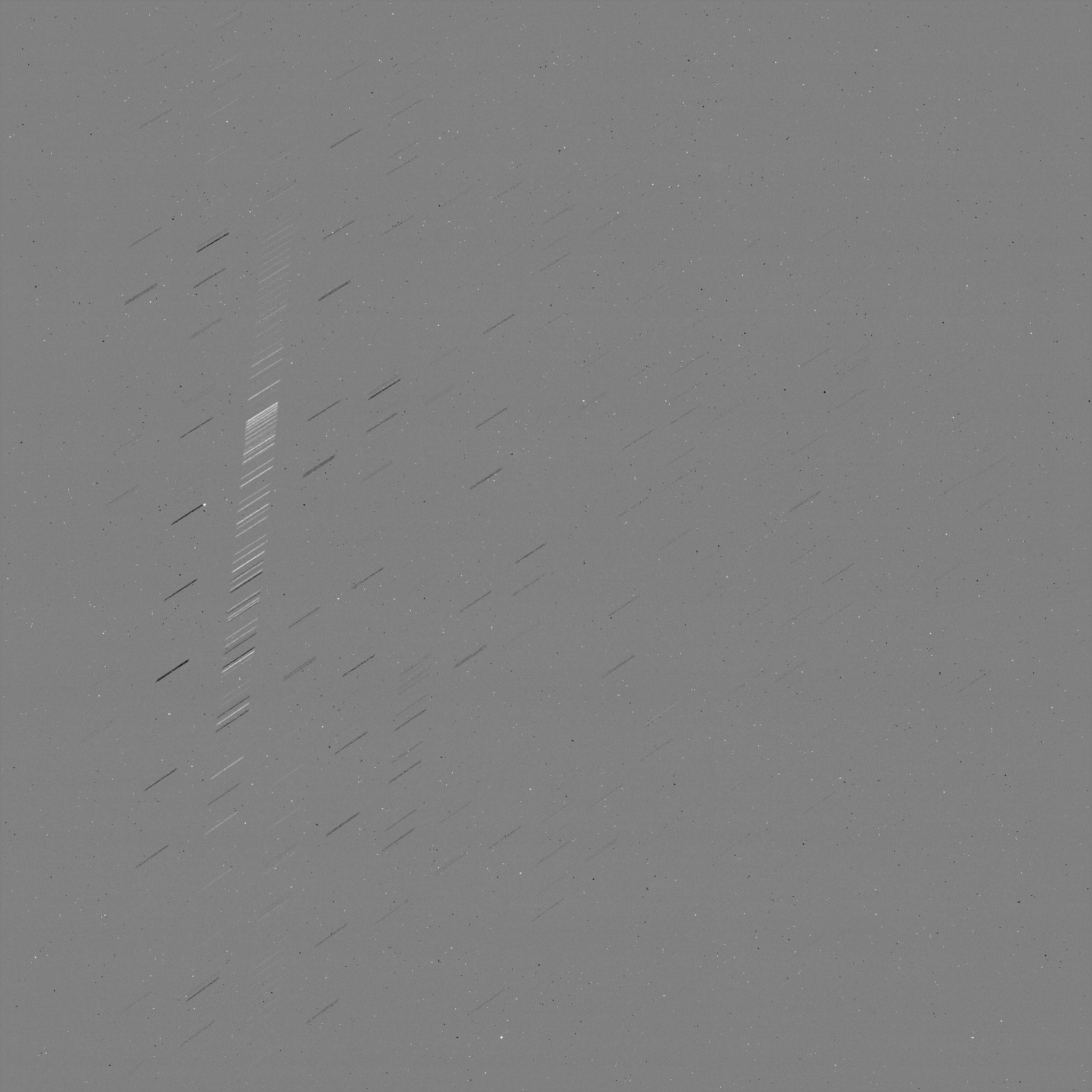}\\
   \includegraphics[width=65mm]{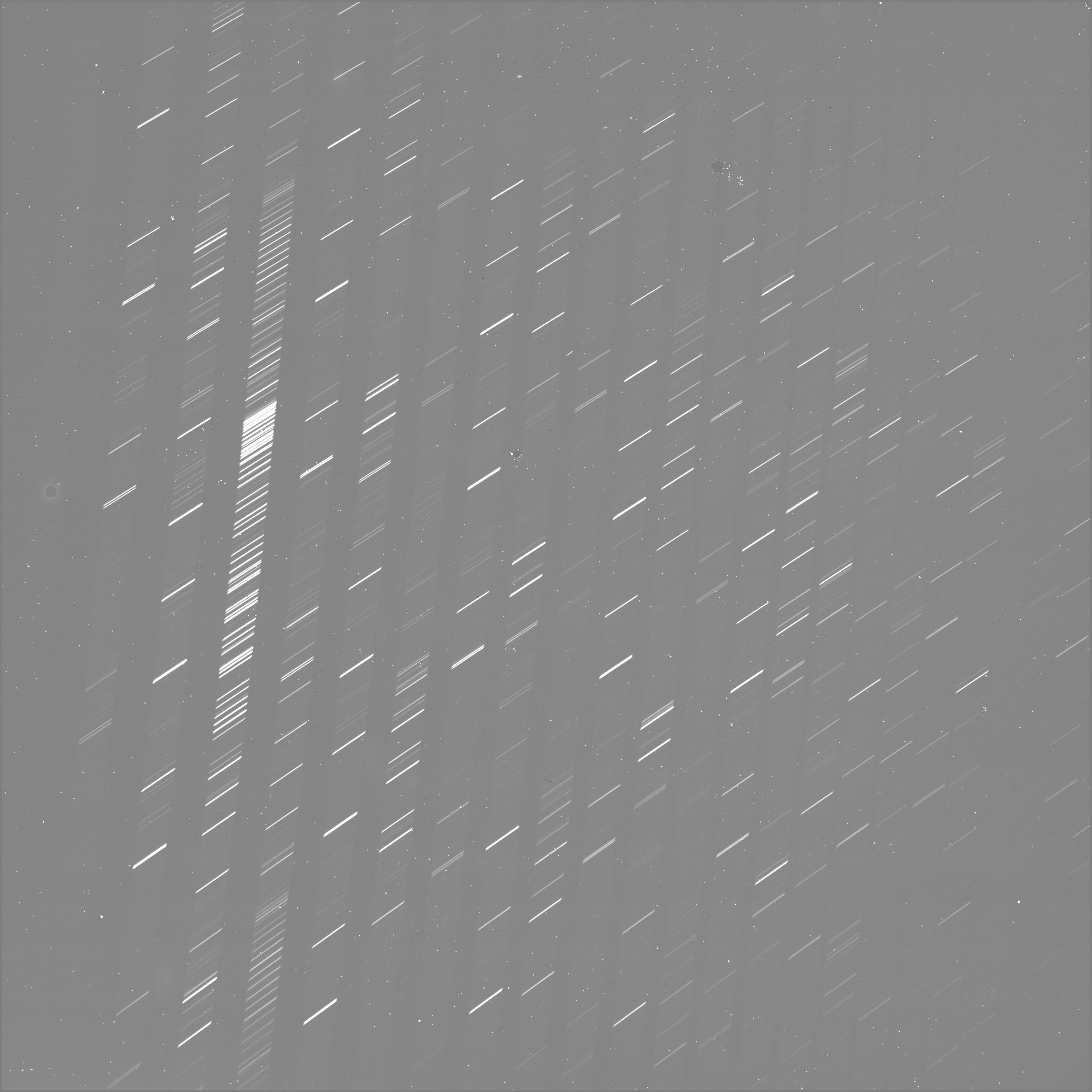}  
    \includegraphics[width=65mm]{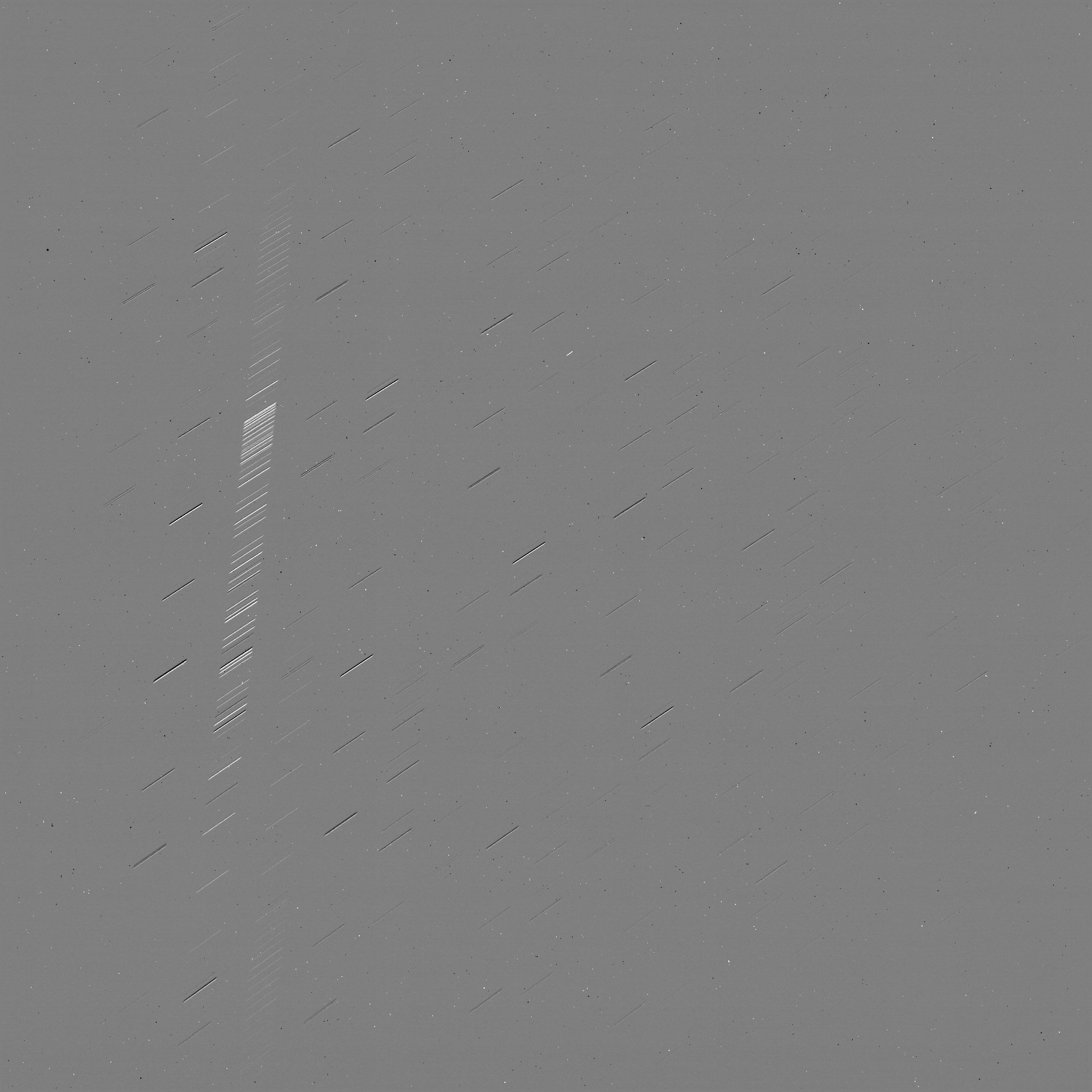}
      \end{center}
\caption{
Examples of slit images before and after the sky subtraction are shown in left and right panels. The upper (lower) panels are for LeoV (Tucana II). The brightness corresponds to the number of photons. The images are reduced to the data of flux/standard deviation in terms of wavelength.   From the 2D position in the figure, we read the wavelength of the photons. We estimate the statistical uncertainty from the photon distribution along the width of the band. 
} \label{fig:slitimage}
\end{figure}

In this supplementary material, we show examples of the slit images before and after sky subtraction in the left and right panels of Fig. \ref{fig:slitimage}, respectively. The top panels are for Leo V and the bottom ones are for Tucana II.

In the right panels, the bright lines are highly suppressed meaning that our sky subtraction is successful, except for the bands whose wavelengths are out of the ``atmospheric window".

The results before subtracting the spline fit of continuous spectra are shown in Fig.\ref{fig:const}. 

\subsection{Limits for generic spectra}
\begin{figure}[!h]
\begin{center}  
   \includegraphics[width=105mm]{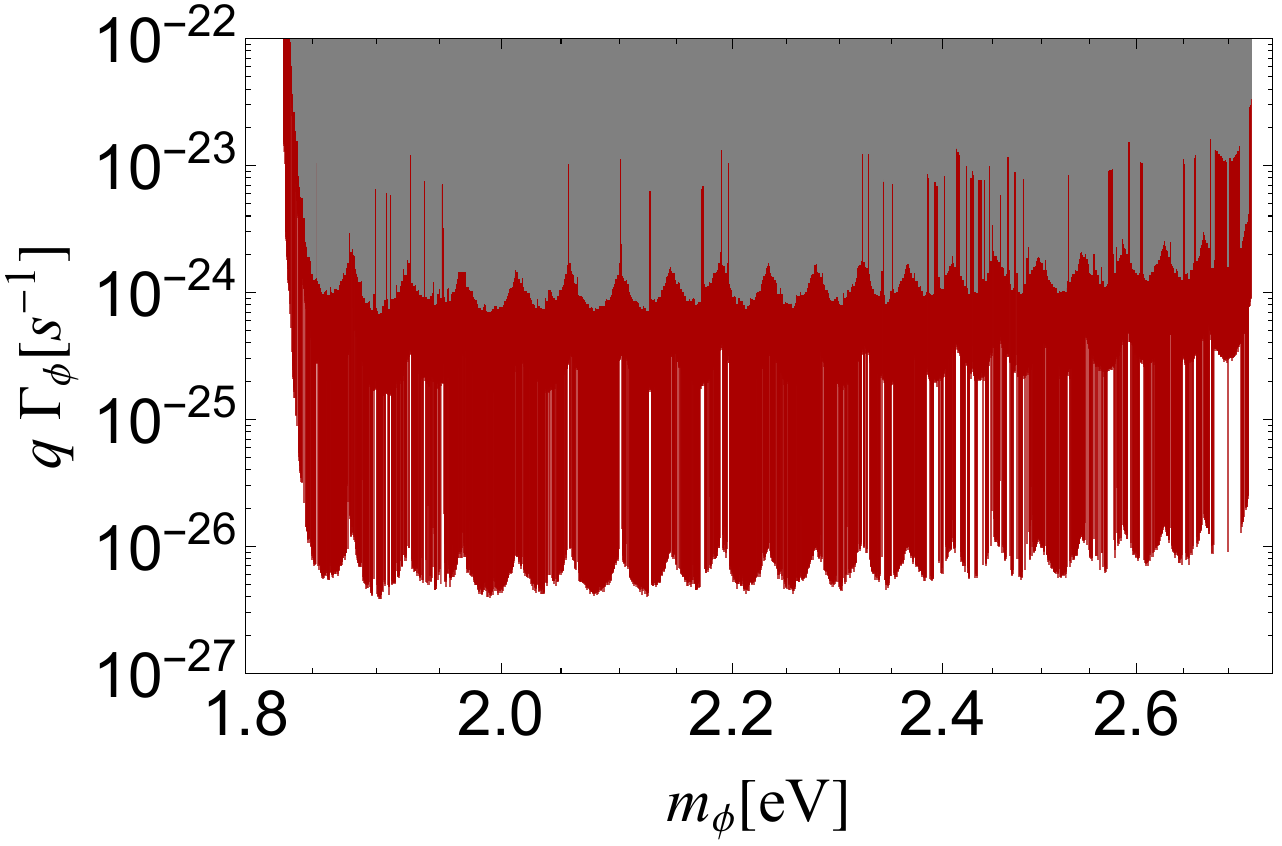}
      \includegraphics[width=105mm]{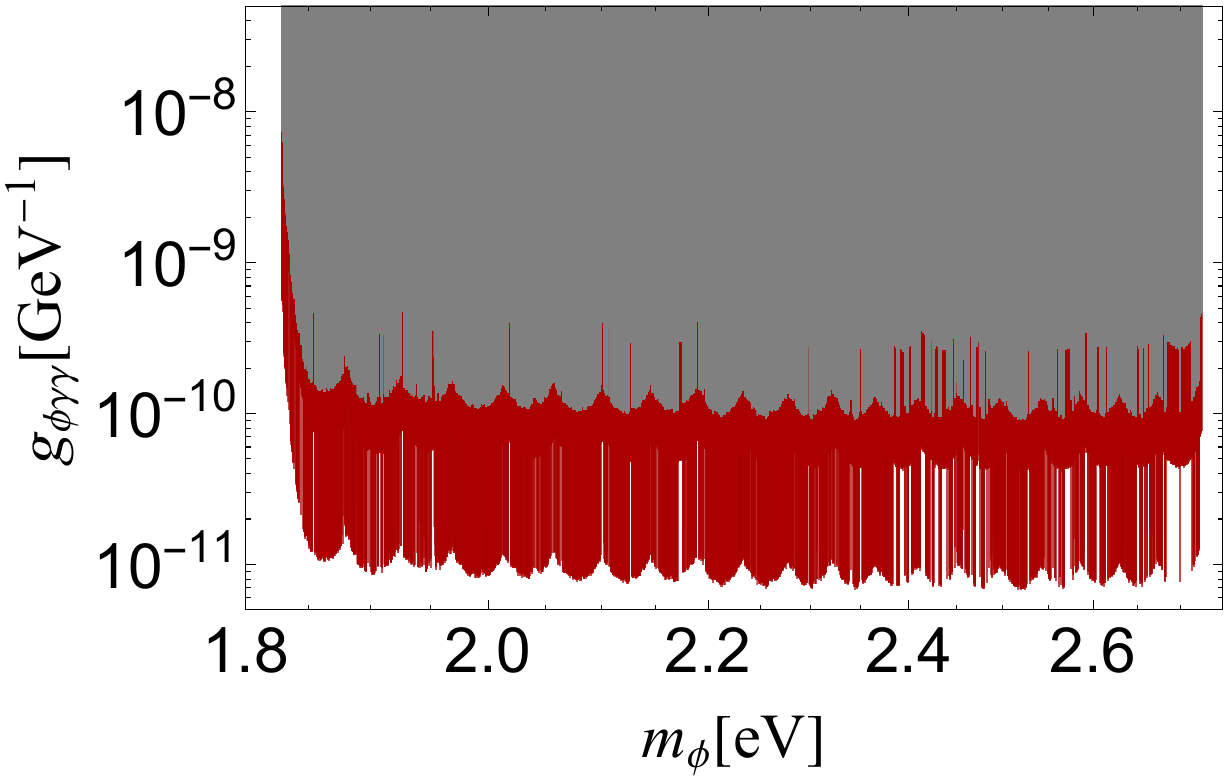}
      \end{center}
\caption{
2$\sigma$ {limit} for dark matter decay into two massless particles, one of which is a photon, before the subtraction of continuous spectrum. 
The top panels show the constraints on the decay rate multiplied by the number of photons in the final state, $q$, by varying the mass $m_\phi$. The red band corresponds to the theoretical uncertainty in the differential $D$-factor assuming a generalized Hernquist profile. 
The bottom panels display the translated constraints on the photon coupling, $g_{\f\g\g}$, for the ALP dark matter ($q=2$). 
} \label{fig:const}
\end{figure}

The {constraint} set in the main part cannot be directly applicable to photons with spectra significantly differing from a line. Such spectra might also be inadvertently subtracted due to the specifics of our background analysis. For those spectra, the data before the background analysis specific for lines are useful. Therefore, we display the result in Fig. \ref{fig:const}.

In Fig. \ref{fig:const}, the shape of the limit is not very smooth, which can be understood as follows.
 The total observation time for Tucana II  is longer than that for Leo V. In addition, the $D$-factor of the Tucana II can be larger than that of LeoV within the uncertainty range. Therefore, the Tucana II data can set a more stringent constraint.
However, there are more Tucana II data to satisfy $|F_i|>2\sigma_i$, and thus, those data are not considered.
 The corresponding data may have been affected by adverse weather conditions on the observation date. In such cases, the data from Leo V prove useful, as they provide {limits} for these bins. 
 
 This {constraint} is less stringent than the one presented in the main part, where we subtract the continuous spectra, primarily because it incorporates less data from Tucana II, but the data used in this figure can be reinterpreted to also constrain photons in different scenarios, such as multi-body decays of dark matter, dark mediator decays~\cite{Jaeckel:2021ert}, or emissions from dark objects in the dSphs. 
They may form continuous spectra.
We can recast our dark matter {limit} to apply to scenarios where the spectra originate from cold dark matter or a fraction thereof. This is because, in such cases, the $D$-factor analysis can also be applied (see, e.g., Ref.\,\cite{Jaeckel:2021ert}).

Since our original data can also be used for different scenarios, which may not relate to the $D$-factor,
we provide the uncertainty of the original data for the zero consistent flux for photon, $2\s_i(\l_{i})/\eta(\l_{i})$ in the attached files. 
Here, the atmospheric transmittance $\eta$ is divided to make the data directly constrain the photon flux out of the atmosphere. 
$\l_i$ is the wavelength on the rest frame at the dSph, $i$. 

When applying a {constraint} derived from a detector with high spectral resolution to line photons of a fixed wavelength, examining the actual data is essential. Subtle features, like vacancies, may not be readily apparent in graphical representations. 
We also provide all the data points for the plots (including the ones in the main part) in the attached files.

\subsection{Differential $D$-factor for the NFW profile}

As the Infrared Camera and Spectrograph (IRCS) and WINERED instruments have excellent angular resolution and small field-of-view, observing near the center can enhance the photon flux, depending on whether the dark matter profile is cuspy, such as the NFW profile~\cite{Bessho:2022yyu,Yin:2023uwf}. Precise estimations of the angular dependence of the photon flux from dark matter decay in 35 dSphs were performed using a generalized Hernquist profile~\cite{Yin:2023uwf}, which we adopted in the main part. However, according to recent $N$-body and hydrodynamical simulations (e.g., \cite{Tollet:2015gqa,Lazar:2020pjs}), the conventional Navarro-Frenk-White (NFW) profile~\cite{Navarro:1995iw, 2010MNRAS.402...21N, Burkert:1995yz} is predicted in ultra-faint dSphs such as Leo V and Tucana II. This led us to also consider the possibility of the NFW profile.

\begin{figure}[!h]
\begin{center}  
   \includegraphics[width=105mm]{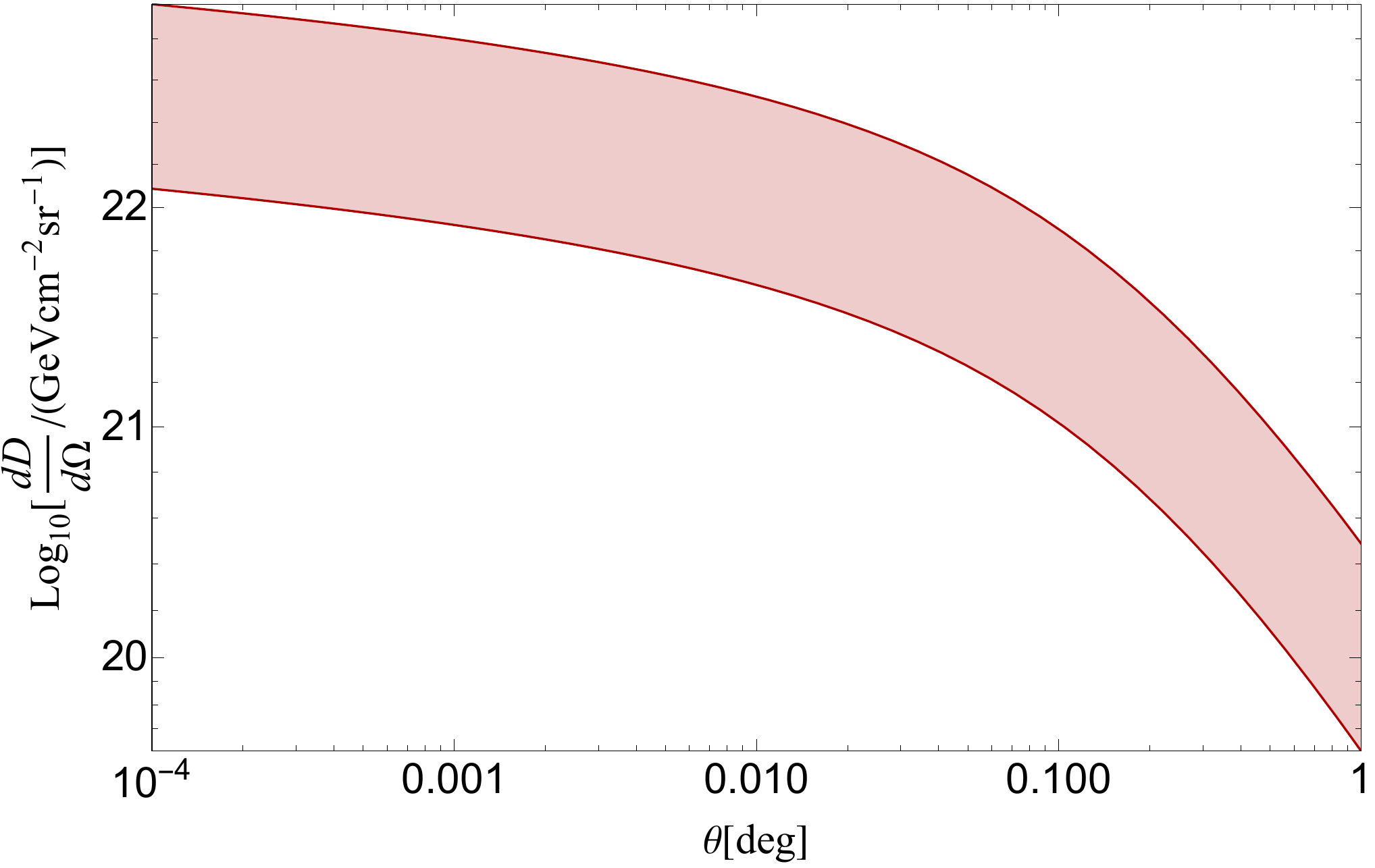}
   \includegraphics[width=105mm]{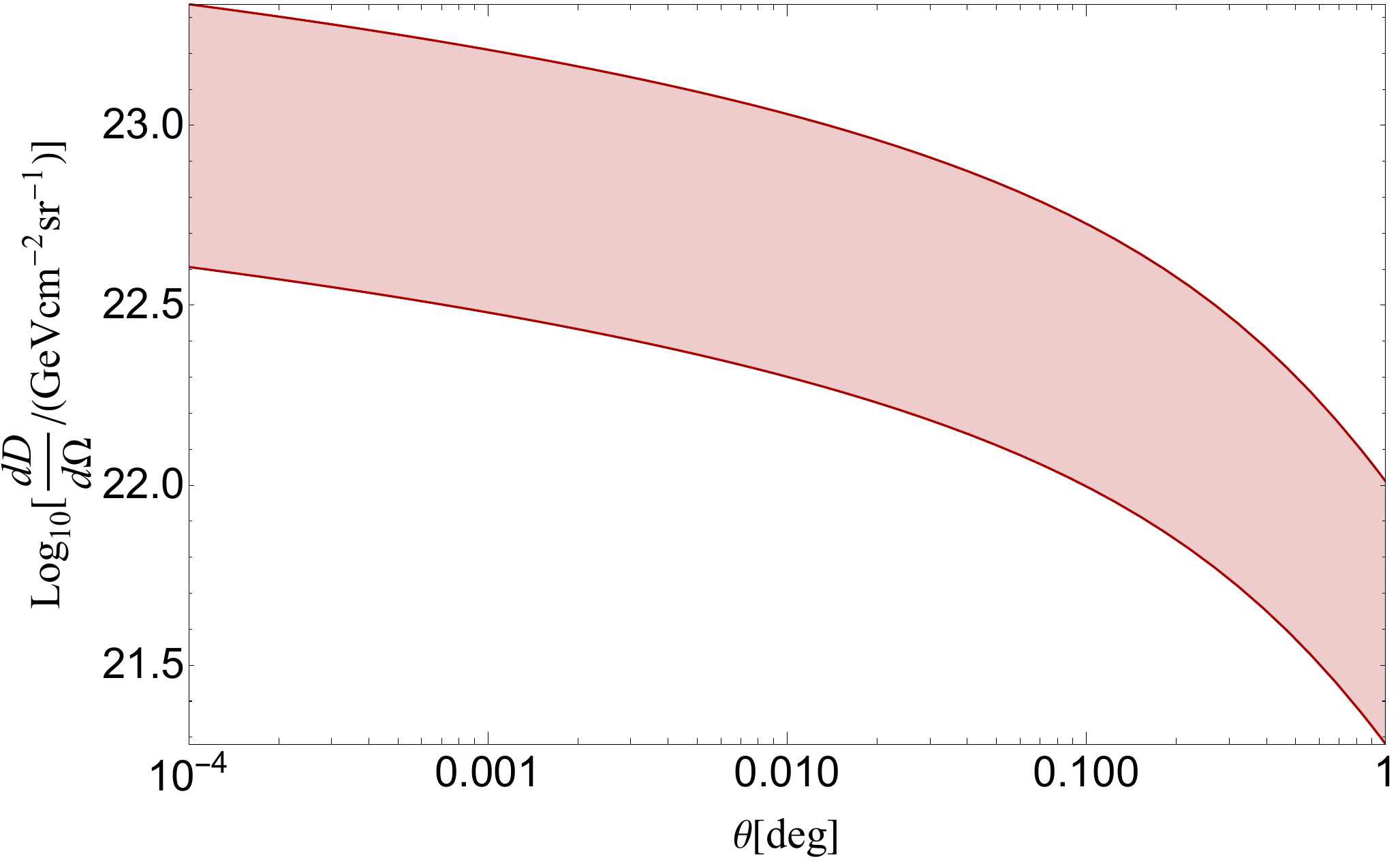}
\end{center}
\caption{
The differential $D$-factor for Leo V (top panel) and Tucana II (bottom panel) assuming an NFW profile.
} \label{fig:NFW}
\end{figure}

Here we recast the $D$-factor for the NFW profile
\beq
\rho_{\rm NFW}(r)= \frac{\rho_0 r_s^3}{r(r+r_s)^2}
\eeq
for the target dSphs, given in Ref.~\cite{Evans:2016xwx}, into the differential $D$-factors shown in Fig.~\ref{fig:NFW}. 
The $D$-factor can be derived as~\cite{Evans:2016xwx}
\beq
D= \frac{4\pi \rho_0 r_s^3 }{d^2} \left[ \log{\left(\frac{d \theta}{2r_s}\right)} + X\left(\frac{d\theta}{r_s}\right)\right],\quad X(s)\equiv \frac{1}{\sqrt{1-s^2}}\, \mathrm{arcsech}[s]\, \Theta(1-s)+\frac{1}{\sqrt{s^2-1}}\, \mathrm{arcsec}[s]\, \Theta(s-1)
\eeq
Here, $d$ is the heliocentric distance, $\theta$ the angular distance from the center of the galaxy and $r_s= 5 R_{\rm half}= 5 d\, \theta_{\rm half}$ with $R_{\rm half}$ being the half-light radius. We use $\theta_{\rm half}=0.07^{\circ}/2$ and $0.5^\circ/2$ for Leo V and Tucana II, respectively, which will give results consistent with the literature. We also mention that the uncertainty of $d$ cancels out and does not enter into the final results in terms of $\theta$. 

By using the result for $D$-factor, $\log_{10}[D^{\rm NFW}_{\rm LeoV}(\theta_{\rm max})/(\GEV \, \mathrm{cm}^{-2})]=16.35^{+0.53} _{-0.35}$ with $\theta_{\rm max}=0.07^\circ$ for Leo V, and $\log_{10}[D_{\rm TucII}^{\rm NFW}(\theta_{\rm max})/(\GEV \, \mathrm{cm}^{-2})]=18.79^{+0.44} _{-0.29}$ with $\theta_{\rm max}=1^\circ$ for Tucana II given in Ref.~\cite{Evans:2016xwx}, we can obtain $\rho_0$ in the dark matter profile with the theoretical uncertainty derived from that of the $D$-factor. 

Then we can estimate 
\beq
\partial_\Omega D\simeq \frac{1}{2\pi \theta}\partial_\theta D[\theta].
\eeq
This is shown in Fig.~\ref{fig:NFW}. Taking $\theta=0.0001^\circ \sim \mathrm{arcsec}$, we get the differential $D$-factor used in the main part, which is the typical position we looked at.

\bibliographystyle{apsrev4-1}
\bibliography{ref}
\end{document}